\documentclass[aps,prl,twocolumn,superscriptaddress,longbibliography]{revtex4-1}

\usepackage{graphicx}
\usepackage{amsmath}
\usepackage{ulem}
\usepackage[usenames,dvipsnames]{xcolor}
\usepackage[colorlinks=true,linkcolor=blue,urlcolor=blue,citecolor=blue]{hyperref}
\usepackage{braket}
\usepackage{amssymb}
\usepackage{commath}
\usepackage{csquotes}
\usepackage{dcolumn}
\usepackage{bm}
\usepackage{epstopdf}
\usepackage{lipsum}
\usepackage{tikz}

\newcommand*\circled[1]{\tikz[baseline=(char.base)]{
		\node[shape=circle,draw,inner sep=0.5pt] (char) {#1};}}

\newcommand{\fig}[1]{Fig.\thinspace{}\ref{#1}}
\newcommand{\fc}[1]{({#1})}
\newcommand{\figc}[2]{Fig.\thinspace{}\ref{#1}\thinspace{}\fc{#2}}

\pdfoutput=1

\begin{document}

\title{Periodically Driving a Many-Body Localized Quantum System}

\author{Pranjal Bordia}
\affiliation{Fakult\"at f\"ur Physik, Ludwig-Maximillians-Universit\"at M\"unchen, Schellingstr. 4, 80799 Munich, Germany}
\affiliation{Max-Planck-Institut f\"ur Quantenoptik, Hans-Kopfermann-Str. 1, 85748 Garching, Germany}

\author{Henrik L\"uschen}
\affiliation{Fakult\"at f\"ur Physik, Ludwig-Maximillians-Universit\"at M\"unchen, Schellingstr. 4, 80799 Munich, Germany}
\affiliation{Max-Planck-Institut f\"ur Quantenoptik, Hans-Kopfermann-Str. 1, 85748 Garching, Germany}

\author{Ulrich Schneider}
\affiliation{Fakult\"at f\"ur Physik, Ludwig-Maximillians-Universit\"at M\"unchen, Schellingstr. 4, 80799 Munich, Germany}
\affiliation{Max-Planck-Institut f\"ur Quantenoptik, Hans-Kopfermann-Str. 1, 85748 Garching, Germany}
\affiliation{Cavendish Laboratory, University of Cambridge, J. J. Thomson Avenue, Cambridge CB3 0HE, United Kingdom}

\author{Michael Knap}
\affiliation{Department of Physics, Walter Schottky Institute, and Institute for Advanced Study, Technical University of Munich, 85748 Garching, Germany}

\author{Immanuel Bloch}
\affiliation{Fakult\"at f\"ur Physik, Ludwig-Maximillians-Universit\"at M\"unchen, Schellingstr. 4, 80799 Munich, Germany}
\affiliation{Max-Planck-Institut f\"ur Quantenoptik, Hans-Kopfermann-Str. 1, 85748 Garching, Germany}

\date{\today}

\begin{abstract}
We experimentally study a periodically driven many-body localized system realized by interacting fermions in a one-dimensional quasi-disordered optical lattice. By preparing the system in a far-from-equilibrium state and monitoring the remains of an imprinted density pattern, we identify a localized phase at high drive frequencies and an ergodic phase at low ones. These two distinct phases are separated by a dynamical phase transition which depends on both the drive frequency and the drive strength. Our observations are quantitatively supported by numerical simulations and are directly connected to the change in the statistical properties of the effective Floquet Hamiltonian. 
\end{abstract}

\maketitle
\paragraph{\textbf{Introduction.---}}Quantum many-body systems far from equilibrium arise naturally in a variety of disciplines, ranging from condensed matter to cosmology. In recent years, there has been an intense focus on understanding the dynamical evolution of quantum many-body systems that are well isolated from their environment~\cite{bloch_many-body_2008, polkovnikov_colloquium_2011}.
Particularly, in periodically driven systems exotic phenomena can emerge that are absent in their undriven counterparts. For example, topologically non-trivial band structures can be realized by driving topologically trivial systems~\cite{oka_photovoltaic_2009, kitagawa_10, Lindner2011,  Aidelsburger13, Miyake13, jotzu_experimental_2014,aidelsburger_measuring_2015} and ergodic phases can be created by driving non-ergodic quantum systems~\cite{ponte_periodically_2015, Ponte15, Achilleas15, abanin_theory_2014, Kozarzewski_Distinctive_2016, Rehn_how_2016, gopalakrishnan_regimes_2016}.

In undriven systems, a robust \textit{non-ergodic} phase can be realized by adding strong disorder to an interacting many-body system, leading to the phenomenon of many-body localization (MBL)~\cite{Anderson58, Basko06, Nandkishore15, Altman15,Kondo15,Schreiber15,smith2015,Bordia16,Hild16}. In an ideal MBL phase, global transport and thermalization are absent, and some memory of the initial conditions persists locally for arbitrarily long times even at finite energy densities~\cite{Nandkishore15,Altman15}, as underlined in experiments~\cite{Schreiber15,smith2015,Bordia16,Hild16}. Recent theoretical works have further proposed that combining MBL and periodic driving can lead to novel symmetry protected topological phases with no direct equilibrium analogues~\cite{PSKhemani16, PSElse16, PS1Keyserlingk16, PS2Keyserlingk16, PSPotter16, FTCElse16, PS3Keyserlingk16}. It is therefore highly pertinent to experimentally study the interplay of disorder and periodic driving in interacting quantum systems.

\onecolumngrid

\begin{figure}[h]
	\centering
	\includegraphics[width=0.98\textwidth]{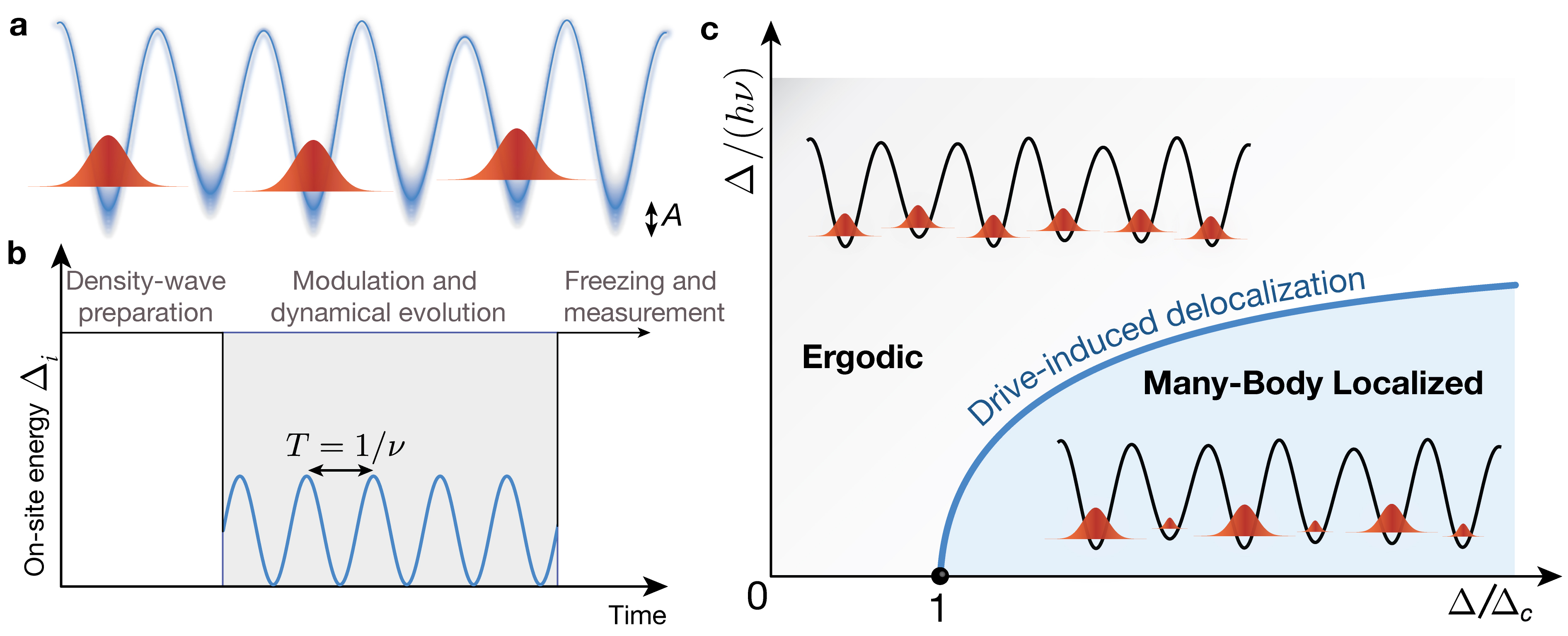}
	\caption{\textbf{Schematic of the experiment and the dynamical phase diagram.} \textbf{(a)} A density-wave pattern of spinful fermionic atoms occupying only the even sites of a disordered optical lattice evolves under \textbf{(b)} a periodic modulation of the on-site disorder potentials $\Delta_i$ with frequency $\nu$ and amplitude $A$. \textbf{(c)} The phase diagram for the strongly driven system ($A = \Delta$) as a function of inverse frequency $1/\nu$ and characteristic disorder strength $\Delta$: In the infinite-frequency limit ($x$-axis), the \textit{disorder-induced} phase transition from an ergodic phase to a many-body localized phase is recovered at a critical disorder strength $\Delta_c$ (black point). While at high but finite drive frequencies the system remains localized for strong disorder, it delocalizes at low drive frequencies. These phases are separated by a \textit{drive-induced} transition (blue line).}
	\label{illustration}
\end{figure}

\twocolumngrid

\begin{figure}[htbp]
	\centering
	\includegraphics[width=.49\textwidth]{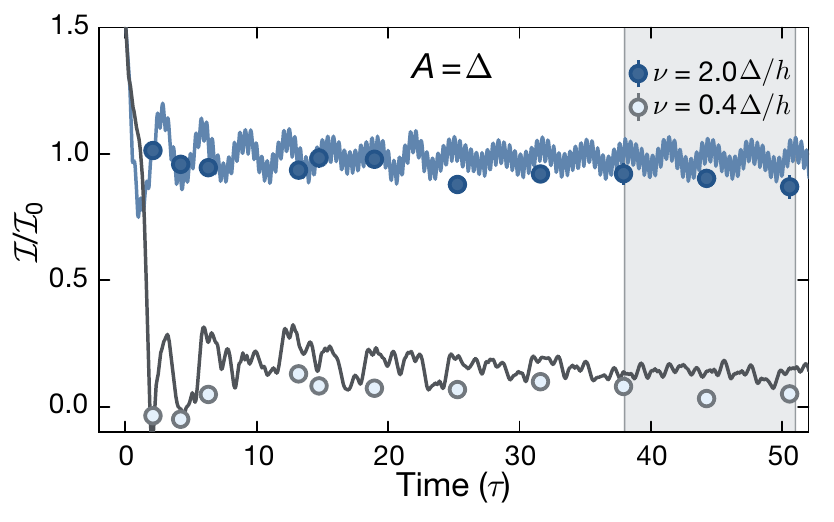}
	\caption{\textbf{Evolution of the imbalance under periodic modulation.} The initially imprinted density-wave pattern, in which atoms occupy only even sites, evolves in a quasi-random disorder potential that is periodically modulated in time. After variable evolution times, we measure the imbalance between the even and odd sites. Fixing the disorder $\Delta = 7.5\,J$ and interaction strength $U = 4\,J$, experimental data (symbols) and numerical simulations (lines) are shown for different values of the drive frequency $\nu$ (legend), for the strongly driven case with $A=\Delta$. Experimental and numerical data are normalized by their respective asymptotic values in the undriven system $\mathcal{I}_0\approx 0.6$~\cite{SOMs}. The experimental data is averaged over six different disorder realizations and the error-of-the-mean (e.o.m) is smaller than the symbol size. The light gray area indicates the time window considered to determine the asymptotic imbalance. Theoretical data to longer times is shown in the supplements~\cite{SOMs}.}
	\label{time_traces}
\end{figure}

In this work, we experimentally study a periodically modulated, disordered many-body system by employing an interacting Fermi gas in a one-dimensional quasi-random optical lattice. The undriven system exhibits a phase transition from an ergodic to an MBL phase as the disorder strength is increased~\cite{Schreiber15}. In presence of a strong drive, we observe a stable MBL phase at high drive frequencies, characterized by a persisting memory on the initial state for long times. In contrast, below a critical frequency, the system delocalizes and completely obliviates the initially imprinted density modulation, see \fig{illustration}. These phases are separated by a dynamical phase transition induced by the periodic \textit{drive}. Our experimental observations are supported by numerical simulations based on matrix product states and exact diagonalization and can be understood as emergent properties of an effective Floquet Hamiltonian.

\paragraph{\textbf{Experiment.---}}Our experimental setup consists of a degenerate $^{40}$K Fermi gas prepared in an equal spin mixture of its lowest two hyperfine states, denoted as $\{\uparrow, \downarrow\}$. We load the gas into the lowest band of a deep three-dimensional optical lattice with at most one atom per site. In the $x$-direction, we employ a superlattice~\cite{Schreiber15} to imprint a density modulation with atoms occupying only even sites of the primary lattice, \figc{illustration}{a}. We initiate the quantum dynamics by lowering the depth of the longitudinal lattice, so that quantum tunneling becomes appreciable along one direction and the density pattern coherently evolves in a one-dimensional quasi-random disorder potential. This quasi-random potential is generated by superimposing a primary lattice beam of wavelength $\lambda_s = 532.0\,$nm with a second laser, the disorder lattice laser, of incommensurate wavelength $\lambda_d = 738.2\,$nm. During the time evolution, we continuously modulate all on-site potentials $\Delta_i$ synchronously by modulating the disorder lattice intensity with frequency $\nu$, \figc{illustration}{b}. After a variable evolution time, we suddenly freeze the system by increasing the depth of the primary lattice and thereby suppressing tunneling. Subsequently, we employ a bandmapping procedure to measure the particle number on even $N_e$ and odd $N_o$ sites in time-of-flight images~\cite{Schreiber15} and calculate the imbalance $\mathcal{I} = (N_e - N_o)/(N_e + N_o)$, which effectively provides a measure for the ergodicity of the quantum system: Under ergodic evolution it rapidly approaches zero while a persistent imbalance indicates non-ergodic dynamics.

\paragraph{\textbf{Model.---}}Our system can be described theoretically by the one-dimensional Aubry-Andr\'e model with on-site interactions~\cite{Iyer13} and a time periodic quasi-random disorder potential:

\begin{align}
\hat{H} =&-J\sum_{i,\sigma}(\hat{c}^{\dagger}_{i+1,\sigma}\hat{c}_{i,\sigma} + \text{h.c.})  +U\sum_{i} \hat{n}_{i,\uparrow}\hat{n}_{i,\downarrow}\nonumber\\&+[\Delta+A \sin (2 \pi \nu t)]\sum_{i,\sigma} \cos (2\pi\beta i+\phi)\hat{n}_{i,\sigma}.
\label{total_hamiltonian}
\end{align}
Here,  $J \approx h \times 550\,$Hz is the tunneling matrix element between neighboring sites and $h$ is the Planck's constant. The fermion creation (annihilation) operator in the spin state $\sigma \in \{\uparrow,\downarrow\}$ on site $i$ are denoted by $\hat{c}^{\dagger}_{i,\sigma}$ ($\hat{c}_{i,\sigma}$) and the particle number operator is $\hat{n}_{i,\sigma}=\hat{c}^{\dagger}_{i,\sigma}\hat{c}_{i,\sigma}$. The on-site interaction strength between the two spin species is given by $U$. The disorder is characterized by the disorder strength $\Delta$, the incommensurate wavelength ratio $\beta = \lambda_s / \lambda_d$, and the relative phase $\phi$.

The disorder is modulated in time with frequency $\nu$ and amplitude $A \in [0,\Delta]$, \figc{illustration}{b}. The total Hamiltonian is thus periodic in time $H(t) = H(t+T)$ with period $T=1/\nu$. For $A=0$ the model has been experimentally shown to exhibit an MBL phase above a critical disorder strength $\Delta_c$ for a wide range of interactions and energy densities~\cite{Schreiber15}. 

\begin{figure*}[t]
	\centering
	\includegraphics[width=.98\textwidth]{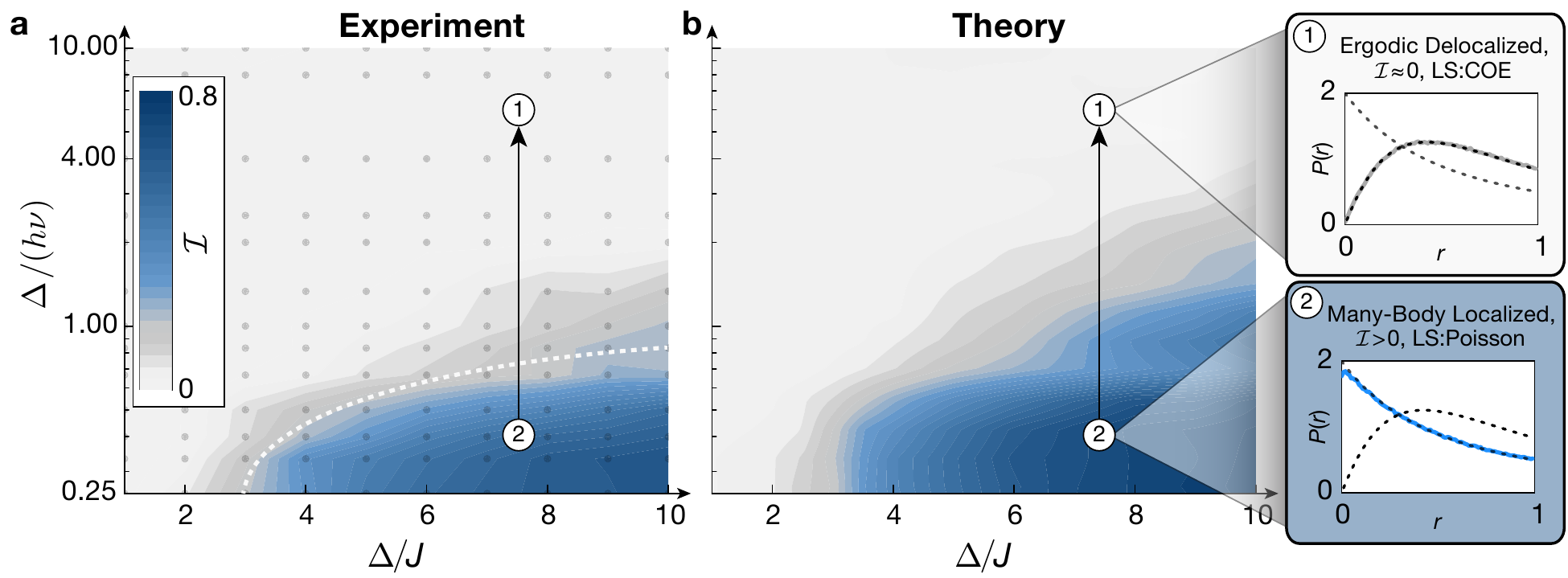}
	\caption{\textbf{Dynamical phase diagram.} Asymptotic imbalance, $\mathcal{I}$, as a function of the disorder strength $\Delta/J$ and inverse frequency $\Delta/\nu$ for strong drive $A=\Delta$ and interactions $U=4\,J$. \textbf{(a)} The interpolated experimental measurements (data taken at the gray dots),  are compared to \textbf{(b)} numerical simulations based on matrix product states. We find a drive-induced delocalization transition (white dashed line, guide to the eye) from the many-body localized phase with high imbalance (blue) to an ergodic phase with vanishing imbalance (gray) when lowering the drive frequency. Using exact diagonalization, the drive induced delocalization transition is also supported by the quasienergy level statistics of the Floquet Hamiltonian, insets to the right. The distribution $P(r)$ of the level statistics parameter $r$ (see main text) follows the circular orthogonal ensemble (COE) in the ergodic phase (gray curve, {\protect\circled{1}}) and Poisson statistics in the localized phase (blue curve, {\protect\circled{2}}). In both insets, the dashed black lines indicate the Poisson and COE distribution functions, respectively.}
	\label{phase_diagram}
\end{figure*}

\paragraph{\textbf{Dynamic Response.---}} First, we concentrate on fixed disorder $\Delta = 7.5\,J$ and interactions $U=4\,J$. For these parameters, the undriven system ($A=0$) is deep in the localized phase and an initial density-wave pattern approaches a stationary state with large residual imbalance $\mathcal{I}_0 \approx 0.6$~\cite{SOMs}. The time evolution of the normalized imbalance $\mathcal{I}/\mathcal{I}_0$ for the strongly driven system $A=\Delta$ is shown in \fig{time_traces} for two different drive frequencies. All times are given in units of tunneling time, $\tau = h/(2\pi J)$. At high drive frequency $\nu = 2\,\Delta/h$ the system remains localized ($\mathcal{I}/\mathcal{I}_0 \sim 1$), implying that the system is transparent to the drive in this regime. In this case, its dynamics is effectively governed by the time averaged non-ergodic Hamiltonian. By contrast, at low drive frequency $\nu = 0.4\,\Delta/h$, the drive enables a redistribution of atoms leading to a vanishingly small imbalance ($\mathcal{I}/\mathcal{I}_0 \sim 0$). Thus, the dynamics become ergodic, even though the time-averaged Hamiltonian is not.
The theoretical data (solid lines), obtained by numerical simulations based on matrix product states, agrees with the experimental measurements (symbols), \fig{time_traces}, and provides strong support for the observed behavior. Furthermore, the good agreement indicates that the system is minimally affected from any external couplings on the experimental time scales~\cite{Bordia16}. 

\paragraph{\textbf{Phase Diagram.---}}To investigate the dynamical phases, we systematically study the long-time asymptotics of the imbalance in the strongly driven system. We measure the asymptotic imbalance $\mathcal{I}$ as the time average between $\sim 40\,\tau-50\,\tau$, as  marked by the light-gray area in \fig{time_traces}. The experimentally measured and theoretically calculated mean imbalance for strong drive $A=\Delta$ and interactions $U=4\,J$ is shown in \fig{phase_diagram} as a function of drive frequency and disorder strength. We note that in this regime, the system is far away from the weak driving considered in linear response. The $x$-axis marks the limit of infinite drive frequency $\Delta/(h\nu) \to 0$, where the imbalance reduces to the one of the undriven system and connects to the phase diagram measured in Ref.~\cite{Schreiber15}.

We observe a stable MBL phase at high but finite frequencies, which is illustrated by the blue area in \fig{phase_diagram}. When lowering the drive frequency at fixed disorder strength and drive amplitude the system undergoes a delocalization transition to the ergodic phase (\raisebox{.5pt}{\textcircled{\raisebox{-.9pt} {2}}} to \raisebox{.5pt}{\textcircled{\raisebox{-.9pt} {1}}}, black arrow). We emphasize that, due to the sinusoidal drive, the fraction of time spent by the system in the delocalized regime is \textit{independent} of the drive frequency. Yet, the nature of the effective dynamics changes completely with its frequency. This arises because at high drive frequencies the atoms only respond to the time-averaged on-site potential, while at low frequencies they can delocalize via the intermediate extended states. The two phases are  separated by a \textit{drive} induced delocalization transition. This phase boundary is indicated by the white dashed line (guide to the eye) in \figc{phase_diagram}{a}, which follows the contour line connecting to the approximate critical disorder strength $\Delta_c \approx 3\,J$ of the undriven system ($x$-axis)~\cite{Schreiber15}. 

For the theoretical interpretation of the observed drive-induced phase transition, it is useful to introduce the Floquet Hamiltonian $\hat H_F$ as
\begin{equation}
 e^{- \frac{i}{\hbar} \, \hat H_F T} = \mathcal{T} e^{-\frac{i}{\hbar} \int_0^T \hat H(t)  dt}
 \label{eq:floquet}
\end{equation}
which describes the unitary evolution over one period of the drive. Here, $\mathcal{T}$ is the time ordering operator and $\hbar$ is the reduced Planck's constant. The Floquet Hamiltonian governs the stroboscopic dynamics of the system and its statistical properties determine whether the quantum dynamics is ergodic or not~\cite{Alessio14}. This can be exemplified by studying the level statistics of the quasienergies $\epsilon_\alpha$ obtained from diagonalizing $\hat H_F$. Due to the periodic drive, the quasienergies $\epsilon_\alpha$ are only defined modulo $h/T$. The distribution $P(r)$ of the level statistics parameter $r_\alpha = \min [\frac{\epsilon_{\alpha+1}-\epsilon_\alpha}{\epsilon_{\alpha}-\epsilon_{\alpha-1}},\frac{\epsilon_{\alpha}-\epsilon_{\alpha-1}}{\epsilon_{\alpha+1}-\epsilon_\alpha}]$ enables one to identify the nature of the phases~\cite{Nandkishore15}. While an ergodic system is expected to follow the circular orthogonal ensemble, a Poisson distribution is expected in the localized phase due to the absence of level repulsion~\cite{Alessio14,Nandkishore15}. Typical plots for the distribution $P(r)$ are shown in the insets in \fig{phase_diagram} and agree with these expectations.
\begin{figure}
	\centering
	\includegraphics[width=.49\textwidth]{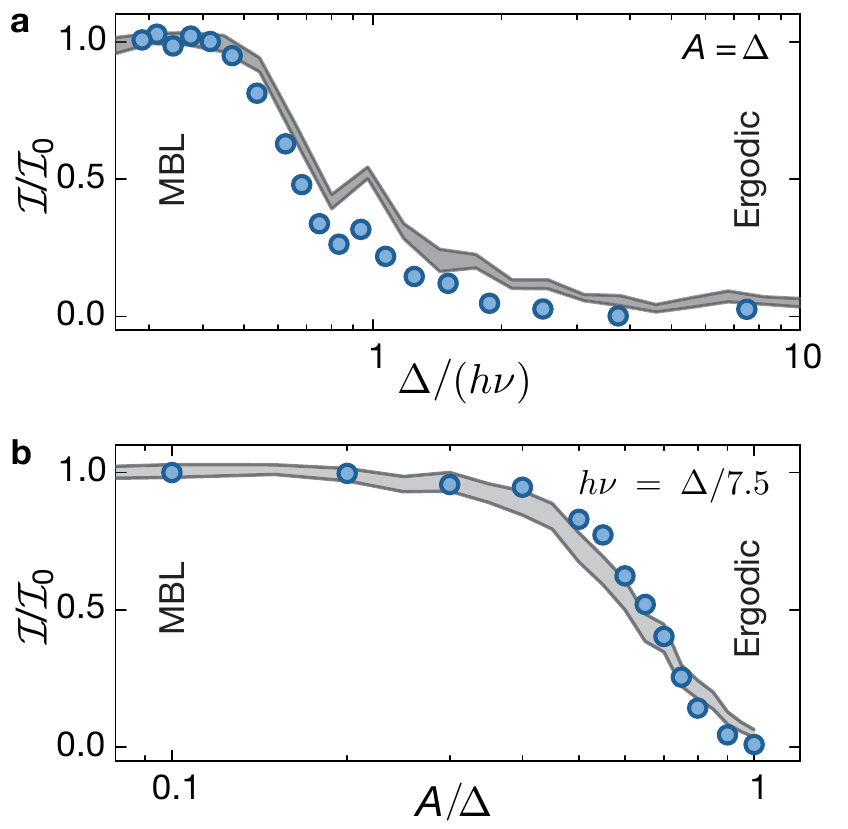}
	\caption{\textbf{Frequency and amplitude dependence of the asymptotic imbalance.} We choose disorder $\Delta = 7.5\,J$ and interaction $U=3\,J$ for which the undriven system is strongly localized. \textbf{\fc{a}} The measured (symbols) and numerical (gray shaded area) data of the asymptotic normalized imbalance $\mathcal{I}/\mathcal{I}_0$ is shown as a function of inverse drive frequency $\Delta/(h\nu)$ for strong drive amplitude $A=\Delta$. \textbf{\fc{b}} Asymptotic imbalance as a function of drive amplitude $A/\Delta$ for comparatively low drive frequency $h\nu=\Delta/7.5$.  While the system remains localized at low amplitudes, it delocalizes for larger ones, with the imbalance continuously approaching zero. E.o.m. over six disorder realizations is smaller than the symbol size. Thickness of the theoretical lines denote one standard deviation of the data for different disorder configurations.}
	\label{cut}
\end{figure}

\paragraph{\textbf{Quantitative Analysis of Frequency and Amplitude Dependence.---}} In order to quantitatively study the frequency dependence of the imbalance for a strong amplitude drive $A=\Delta$, we choose disorder $\Delta = 7.5\,J$ and interactions $U=3\,J$ and illustrate in \figc{cut}{a} the transition from the localized phase at high frequencies to the ergodic phase at low ones. The imbalance decreases continuously with decreasing frequency except for a small peak around $\Delta/(h\nu) \approx 1$, a fact which is also discernible in theoretical contours in \fig{phase_diagram} and stems from the energy level distribution of the quasi-periodic disorder potential~\cite{SOMs}. At low frequencies with $\Delta/(h\nu) \gtrsim 2$, we measure close to vanishing imbalances signaling ergodic dynamics. In this regime, small residual imbalance in the theoretical simulations follows a slow power law relaxation~\cite{SOMs}. Due to the finite lifetime in the experiment, the theoretically evaluated imbalance is typically slightly larger than the experimental one~\cite{Bordia16}.

The response of the system as a function of the drive amplitude at low drive frequency $h\nu=\Delta/7.5$ is shown in \figc{cut}{b}. The normalized imbalance decreases as a function of the drive amplitude and for strong drives ($A\approx\Delta$), the system delocalizes with vanishing steady state imbalance. Crucially, the system appears to remain \textit{localized} for small amplitudes even though the drive frequency is small. This seems not to be a finite time effect but rather a consequence of the drive modulating all on-site energies synchronously, i.e., modulating the disorder strength itself. At low enough drive frequency, weakly modulating the disorder strength results in a non-equilibrium state in which the localization length, taken as the spatial extent of the local conserved quantities~\cite{Nandkishore15}, is modulated adiabatically in time. Such a modulation cannot generate enough resonances to destabilize the system and hence the imbalance remains finite. Following this argument, interactions only lead to a dephasing of the conserved quantities and do not facilitate the redistribution of the atoms. 
This result suggests a novel periodically driven regime which would not delocalize for \textit{any} frequency, as supported by numerical simulations to long times (see supplements~\cite{SOMs}). We note that this apparent stability at low frequencies does not contradict previous theoretical studies, which have focused  on oscillating linear potentials~\cite{abanin_theory_2014, gopalakrishnan_regimes_2016, Kozarzewski_Distinctive_2016}. In that case, the system can always delocalize at low enough frequencies because of long-distance Landau-Zener crossings~\cite{abanin_theory_2014, gopalakrishnan_regimes_2016}. In the supplements, we also report the data for the full dynamical phase diagram as a function of the amplitude and frequency~\cite{SOMs}.

\paragraph{\textbf{Conclusions and Outlook.---}}By periodically driving a many-body localized system, we have created and observed non-ergodic and ergodic phases that emerge from an effective Floquet Hamiltonian and are separated by a drive-induced delocalization transition. Our results directly show dynamics beyond the time-averaged Hamiltonian. Periodic modulation paves the way for measuring the frequency resolved response of many-body systems, with the particular prospect of exploring the critical point of the MBL transition~\cite{Agarwal15, Vosk_Theory_2015, Gopalkrishnan15} and the response in higher dimensions~\cite{Hild16}. Furthermore, our observations demonstrate that disorder can be used to protect interacting and periodically driven systems from heating to infinite temperature and hence sets the basis for realizing novel symmetry protected topological phases~\cite{PSKhemani16, PSElse16, PS1Keyserlingk16, PS2Keyserlingk16, PSPotter16, FTCElse16, PS3Keyserlingk16}.\

\paragraph{\textbf{Acknowledgments.---}}We thank E. Altman, E. Demler, S. Gopalakrishnan, and S. Hodgman for many useful discussions. We acknowledge support from Technical University of Munich - Institute for Advanced Study, funded by the German Excellence Initiative and the European Union FP7 under grant agreement 291763, the European Commission (UQUAM, AQuS) and the Nanosystems Initiative Munich (NIM).

\newpage

\cleardoublepage
	
\bibliographystyle{nphys}
\bibliography{FloquetMBL_arxiv1}

\cleardoublepage

\appendix

\setcounter{figure}{0}
\setcounter{equation}{0}

\renewcommand{\thepage}{S\arabic{page}} 
\renewcommand{\thesection}{S\arabic{section}} 
\renewcommand{\thetable}{S\arabic{table}}  
\renewcommand{\thefigure}{S\arabic{figure}} 

\section{\Large{Supporting Material}}
\setlength{\intextsep}{0.8cm} 
\setlength{\textfloatsep}{0.8cm}

\paragraph{\textbf{Modulating the disorder strength.---}}To realize the periodic drive, we  sinusoidally modulate the intensity of the laser that realizes the disorder lattice. The intensity of this laser is linearly proportional to the onsite disorder and hence modulates the onsite disorder in a sinusoidal fashion~\cite{Schreiber15}.\\

\begin{figure}[b!]
	\centering
	\includegraphics[width=01.0\linewidth]{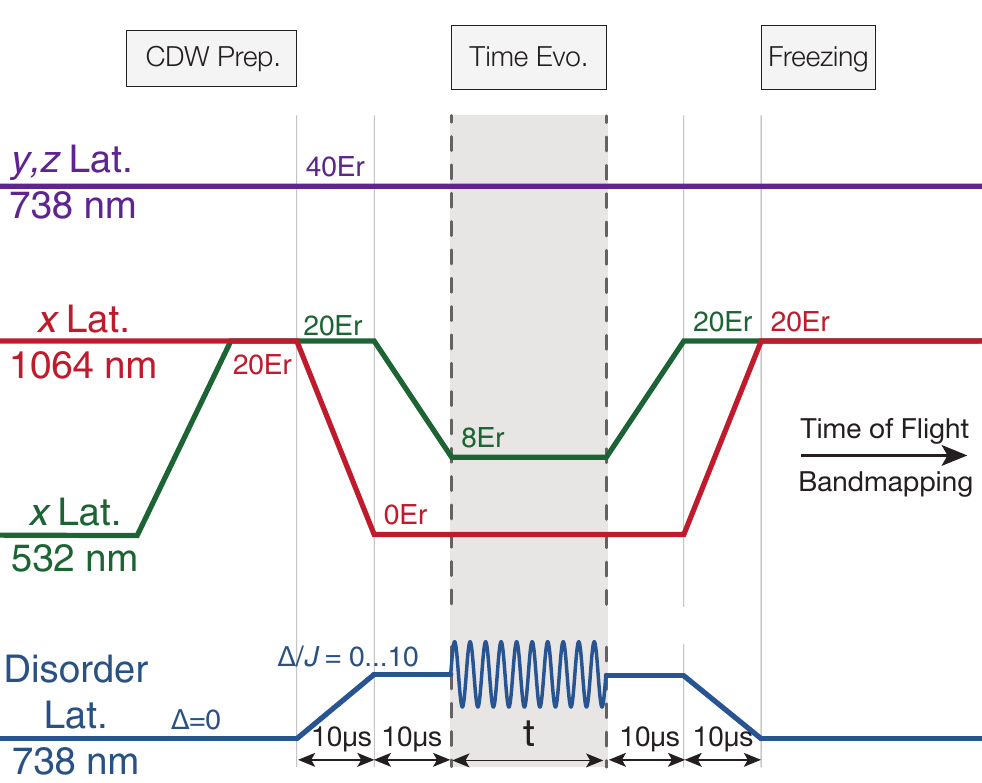}
	\caption{\textbf{Lattice sequence for the experiment.} In order to keep the system effectively one-dimensional we freeze the motion along the perpendicular directions by ramping the $y,z$-direction lattices deep to 40$E_{\text{r}}$. Along the $x$-direction we prepare a density-wave  pattern using the superlattice where only the even sublattice is occupied. We then initiate the dynamics by switching off the long lattice, and quickly ramping the short lattice to 8$\,E_{\text{r}}$. Simultaneously, we modulate the disorder lattice at a given frequency and amplitude. Afterwards, we freeze the system by suddenly ramping up the lattices and initiate the bandmapping procedure to extract the total number of atoms on even and odd sublattices. Note that the $x$-axis is not to scale.}
	\label{sequenceprepdet}
\end{figure}

\paragraph{\textbf{General experimental sequence.---}}We produce a gas of ultracold fermionic Potassium-40 ($^{40}$K) atoms by sympathetically cooling it with bosonic Rubidium-87 ($^{87}$Rb). By lowering the trapping potential depth, we let the gravitational force selectively remove the $^{87}$Rb atoms, leaving behind a pure fermionic gas. This Fermi gas consists of an equal mixture of the two spin components. We evaporate the gas further and obtain a final temperature of $0.15(2)\,T_F$, where $T_F$ is the Fermi temperature. The two spin components are the two lowest hyperfine states of the $^4S_{1/2}$  manifold ($\ket{F,m_F} = \ket{9/2,-9/2} \equiv \ket{\downarrow}$ and $\ket{9/2,-7/2} \equiv \ket{\uparrow}$). Interactions between the two spin components are tuned via an $s$-wave Feshbach resonance at 202.1G~\cite{Regal03}.

We load the gas in to a deep three dimensional lattice where tunneling can be neglected. The lattice is formed by lasers of wavelengths $\lambda_{\text{l}} = 1064\,$nm in the $x$-direction and  $\lambda_{\text{y,z}} = 738.20\,$nm in the perpendicular directions. In order to avoid double occupancies in the initial state, the interactions between the two species are set to be repulsive during loading with a magnitude of 140$\,$a$_0$, where a$_0$ is the Bohr radius. This restricts the double occupancies in the initial state to less than $10\%$. Next, we prepare a density wave by utilizing a superlattice setup in the $x$-direction~\cite{Trotzky08}, consisting of a short $\lambda_{\text{s}}=532.0\,$nm lattice laser in addition to the long lattice laser $\lambda_{\text{l}}$ . Subsequently, we set the interactions to the desired value using the Feshbach resonance and initiate the dynamics by lowering the $x$-lattice to $8\,E_{\text{r}}^{\text{s}}$~\cite{Schreiber15}. The system remains effectively one-dimensional during the time scales probed in the experiment as quantum tunneling is appreciable only in the $x$-direction. The perpendicular lattices remain deep at $40\,E_{\text{r}}^{\text{y,z}}$ and hence quantum tunneling is strongly suppressed. Here $E_{\text{r}}^i = h^2/(2 m \lambda_i ^2)$ is the recoil energy of the laser with wavelength $\lambda_i$ and $m$ is the atomic mass. 

Simultaneous to the reduction of the $x$-lattice depth, we introduce quasi-random disorder by illuminating the atoms with a second, incommensurate lattice with $\lambda_d = 738\,\text{nm}$, which we refer to as the disorder lattice. By modulating the intensity of the disorder lattice, we synchronously modulate all onsite disorder potentials, see Eq.(1). Different phases $\phi$ of the disorder are realized by changing the detuning of the disorder laser. After variable evolution times, we suddenly ramp up the short lattice within 10$\,\mu \text{s}$ to 20$\,E_{\text{r}}^{\text{s}}$ followed by ramping the long lattice to 20$\,E_\text{r}^{\text{l}}$ to suppress the tunneling. Then we initiate our bandmapping sequence to measure the population of even and odd sublattices and perform absorption imaging on the bandmapped atomic gas after 8$\,$ms of time-of-flight~\cite{Trotzky12,Schreiber15}. A brief schematic of this lattice sequence is given in \fig{sequenceprepdet}.\\

\begin{figure*}
	\centering
	\includegraphics[width=1.0\linewidth]{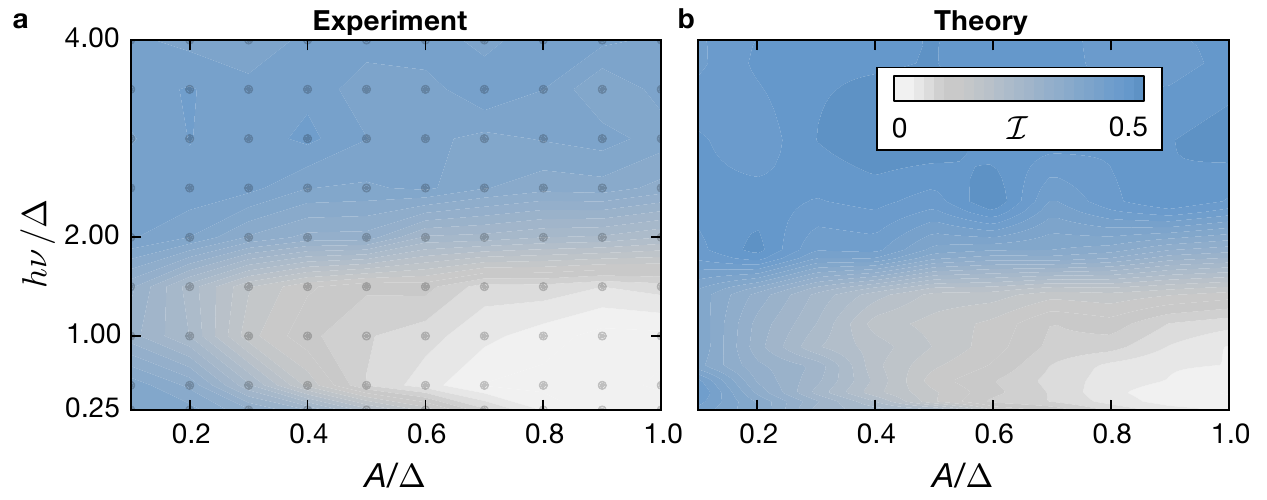}
	\caption{\textbf{Dynamical phase diagram as a function of frequency and amplitude.} The asymptotic imbalance is presented for disorder strength $\Delta = 4\,J$ and interaction strength $U = 4\,J$. We observe that the system remains localized except for large drive amplitudes and low frequencies. In the experimental graph, the gray dots denote the parameters where the data is measured.}
	\label{pdampnu}
\end{figure*}

\paragraph{\textbf{Atom losses and band resonances.---}} We measure no discernible atom loss with or without shaking for all measurements shown in the main text. This is because higher band resonances are off-resonant in energy and hence transitions to higher bands are suppressed. Due to the parity of the Wannier functions, the lowest band to which the atom could be excited via amplitude modulation is the third band (second excited band) along the $x$-direction which starts with a gap of $210\,J$ above the lowest band. This is much larger than the largest frequency probed in the experiment of $40\,J$. In addition, the background lifetimes of the imbalance and atom number are very long compared to the time scales probed in the main text~\cite{Bordia16}.\\

\paragraph{\textbf{Amplitude-frequency phase diagram.---}} The full phase diagram at $\Delta = 4\,J$ and $U = 4\,J$ is shown in \fig{pdampnu} as a function of amplitude and frequency of the drive for both experiment and theory. We observe that the system remains localized except for large amplitudes ($A \sim \Delta$) and low frequencies.\\

\paragraph{\textbf{SU(2) symmetry.---}}In the experiment, a weak breaking of the SU(2) symmetry between the spin states is expected due to the different polarizabilities of the two hyperfine states at finite detuning of the lasers. For the numerical studies, we assume a 2$\%$ difference in the disorder strength for the two spin species which breaks the SU(2) symmetry.\\

\paragraph{\textbf{Asymptotic imbalance value $\mathcal{I}_0$ in the undriven system.---}}The asymptotic values of the undriven system, $\mathcal{I}_0$, is also obtained from such time traces both for theory and experiment. For disorder $\Delta = 7.5\, J$ and $U = 4\,J$, the undriven asymptotic value measured in the experiment is $\mathcal{I}_0^{\text{expt}} = 0.56$ and for the theory is $\mathcal{I}_0^{\text{theory}} = 0.66$. The difference between theory and experiment is primarily attributed to the finite background lifetime of the experiment \cite{Bordia16} and to the fidelity of the initial state imbalance of $\mathcal{I}_\text{expt}(t=0) = 0.91$, compared to the numerical calculations which has unity imbalance.\\

\paragraph{\textbf{Numerical simulations.---}}We employ matrix product states~\cite{vidal_efficient_2004} to numerically investigate the time-dependent imbalance of the periodically modulated interacting Aubry-Andr\'e model. In all the presented data, we choose systems of size $L=32$. We discretize the time evolution operator using a second order Trotter decomposition with timestep $0.05\, \tau$ and take a bond dimension of 800 for the matrix product states. This guarantees that the truncated weight remains small; typically smaller than $5 \times 10^{-6}$. For the initial state, we prepare a density-wave pattern in which every other site is occupied by a fermion in a random spin state. For each time trace of the imbalance we average over $20-40$ disorder realizations. The asymptotic imbalance shown in the phase diagrams is obtained by averaging the traces between 30$\, \tau$ and 50$\, \tau$, which is based on a slightly larger time window compared to the experimentally determined value, in order to remove temporal oscillations in the imbalance.

\begin{figure}[h]
	\centering
	\includegraphics[width=1\linewidth]{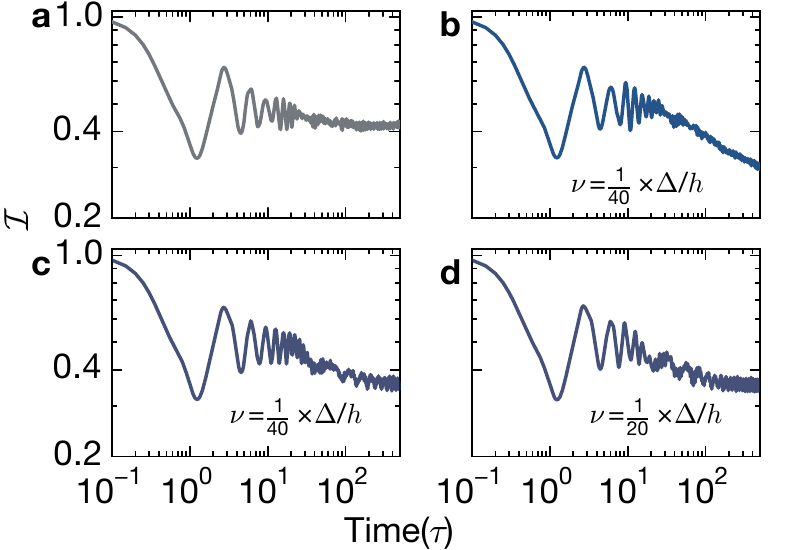}
	\caption{\textbf{Long-time dynamics of the imbalance for different driving modes.} For disorder $\Delta=5\,J$ and interactions $U=4\,J$, we calculate the time trace of the imbalance up to times $500 \, \tau$. We compare the dynamics without the drive $A=0$ \textbf{(a)}, where we observe the imbalance to saturate at a finite value, to finite amplitude drive $A=0.1\, \Delta$ for different driving modes \textbf{(b--d)}. While a field gradient modulation with drive frequency $\nu=\frac{1}{40}\times\Delta/h$ results in a power law decay in the imbalance \textbf{(b)}, a globally homogeneous modulation of the on-site energies shown for \textbf{(c)} $\nu = \frac{1}{40}\times\Delta/h$ and \textbf{(d)} $\nu = \frac{1}{20}\times\Delta/h$, indicates slow relaxation to a finite imbalance.}
	\label{gradientamplitude}
\end{figure}

We compute the level statistics of the effective Floquet Hamiltonian by approximating the sinusoidal drive with a two step function. Using exact diagonalization, we obtain all the eigenstates and eigenenergies of the stroboscopic time evolution operator for systems of $L=8$ sites at quarter filling. This allows us to compute the level statistics $r$ for the quasi-energies of the Floquet Hamiltonian. The level statistics distribution $P(r)$ shown in the main text consists of data from 2000 disorder realizations.\\

\begin{figure}
	\centering
	\includegraphics[width=1.0\linewidth]{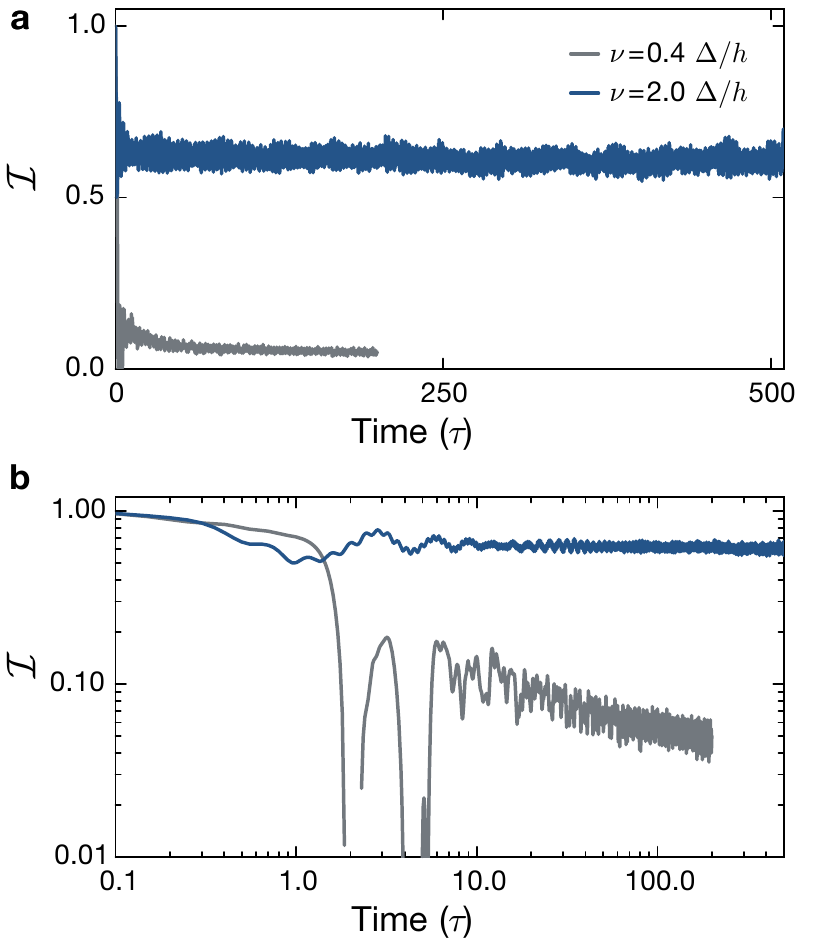}
	\caption{\textbf{Long-time dynamics of large amplitude drive}. Evolution of the imbalance for disorder $\Delta = 7.5\, J$ and interaction $U = 4\, J$ with amplitude $A= \Delta$, as in \fig{time_traces} but for longer times supporting the observation that the system is localized for high frequency $\nu = 2.0\,\Delta/h$ but delocalizes for low frequency $\nu = 0.4\,\Delta/h$. Part \textbf{(a)} and \textbf{(b)} show the same data but on a linear-linear and log-log plot, respectively. While \textbf{(a)} allows direct comparison with \fig{time_traces}, \textbf{(b)} demonstrates the slow power law decay of small residual imbalance at low drive frequency of $\nu = 0.4 \, \Delta/h$.}
	\label{tt_long}
\end{figure}

\paragraph{\textbf{Long-time dynamics of weakly driven system.---}} We numerically simulate the long time dynamics of the imbalance for different driving modes and particularly focus on the case of weak drive amplitude $A \ll \Delta$ and small frequencies $\nu\ll \Delta/h$, \fig{gradientamplitude}. We set the disorder to $\Delta = 5\,J$ and the interaction strength to $U = 4\,J$. We compare the relaxation of the \textbf{(a)} undriven system, where we observe saturation at a finite imbalance, to \textbf{(b--d)} weakly driven systems $A=0.1\,\Delta$ with different driving modes. For this simulation, the starting state is a product state with alternating spin states (unlike the rest of the paper which has random configuration of initial spins states). This fixed initial state has been chosen to remove the uncertainty resulting from sampling different spin configurations and hence reduces the number of samples required to obtain a converged time trace. The modulation of a field gradient, corresponding to a linearly increasing potential in space, leads to a power-law decrease of the imbalance at late times \textbf{\fc{b}}. In contrast, a globally homogeneous drive that synchronously modulates all on-site potentials, leads to a saturation of the imbalance to a finite value. This is shown for two small frequencies in subplot \textbf{(c)} for $\nu = \frac{1}{40}\times\Delta/h$, and in \textbf{(d)} for $\nu=\frac{1}{20}\times\Delta/h$. Such behavior may be explained by the fact that the system follows the drive quasi-adiabatically. While this leads to a periodic modulation of the localization length, set by the spatial extent of the locally conserved quantities~\cite{Nandkishore15}, this drive seems not to be able to create a percolating network of resonances and therefore does not lead to global particle rearrangements, which renders a persistent localization plausible. Studying such extreme long-time dynamics remains challenging for the current experimental setup due to the residual background effects, such as coupling to the neighboring tubes~\cite{Bordia16}.\\

\begin{figure}
	\centering
	\includegraphics[width=1\linewidth]{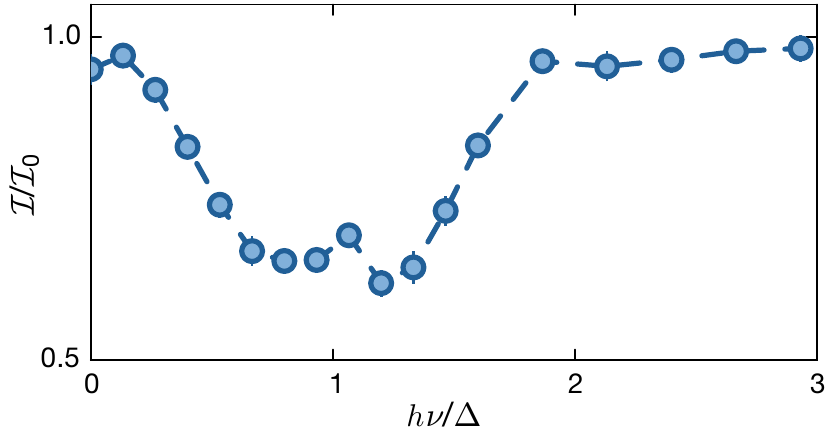}
	\caption{\textbf{Frequency resolved spectrum.} The figure shows normalized imbalance for the non-interacting system with disorder $\Delta = 7.5\,J$ as a function of drive frequency $\nu$ at amplitude $A=0.2\,\Delta$.}
	\label{nuSpectrum}
\end{figure}

\begin{figure}
	\centering
	\includegraphics[width=1\linewidth]{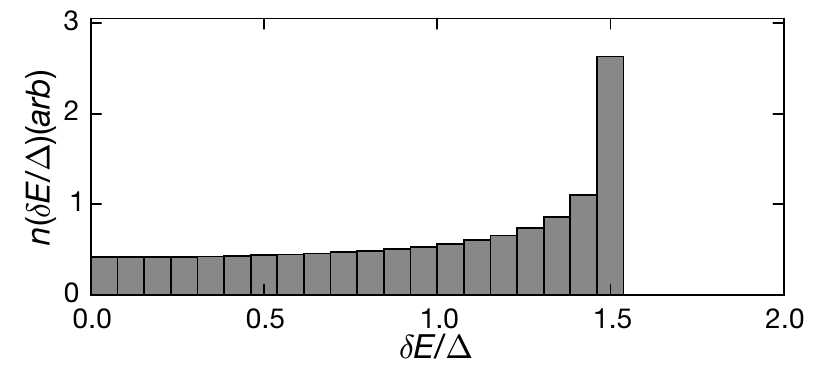}
	\caption{\textbf{Nearest-neighbor energy differences for quasi-random disorder.} The energy difference of nearest-neighbor on-site energies realized by quasi-random disorder of the Aubry-Andr\'e model is peaked at around $1.5\,\Delta$ which gives rise to the small substructure in the frequency resolved spectrum.}
	\label{aaediff}
\end{figure}

\paragraph{\textbf{Long-time dynamics of the strongly driven system.---}}We compute the long-time dynamics corresponding to \fig{time_traces}, which is shown in \fig{tt_long}. These results support the observation that the system is localized at large frequency $\nu = 2.0\,\Delta/h$ while it is delocalized at small frequency $\nu = 0.4\,\Delta/h$. For strong drive and small frequency, the numerical data suggests that after a fast initial drop, the residual imbalance decays as a powerlaw toward zero with a small exponent, while it rapidly saturates for high drive frequencies.\\

\paragraph{\textbf{Frequency resolved absorption.---}}In order to determine the absorption spectrum of the system, we measure the imbalance as a function of the drive frequency for relatively small drive amplitude $A = 0.2\,\Delta$, \fig{nuSpectrum}. We consider strong disorder $\Delta = 7.5\,J$ and set the interactions close to $U=0$. The absorption spectrum features a reduced imbalance for frequencies between 0.2 and 2$\,\Delta/h$. We understand the data as follows: At low frequencies the system can adiabatically follow the drive and hence the imbalance remains high. Upon increasing the frequency the system begins to absorb from the drive. However, for even larger frequency, the system cannot follow the drive anymore and only evolves with the time-averaged Hamiltonian. Hence the system effectively becomes transparent to the drive and remains localized. We attribute the peculiar line shape of a double dip, to the very asymmetric distribution of nearest-neighbor onsite energy differences in the quasi-periodic Aubry-Andr\'e disorder potential, which features a peak at $1.5\,\Delta$, see \fig{aaediff}. This strongly peaked distribution also gives rise to the relatively sharp features visible in the simulations shown in \fig{phase_diagram} and~\figc{cut}{a}. In the experiment, the resulting structure might be softened due to the presence of the harmonic confinement and averaging over multiple 1D tubes~\cite{Schreiber15}.\\

\begin{figure}
	\centering
	\includegraphics[width=1.0\linewidth]{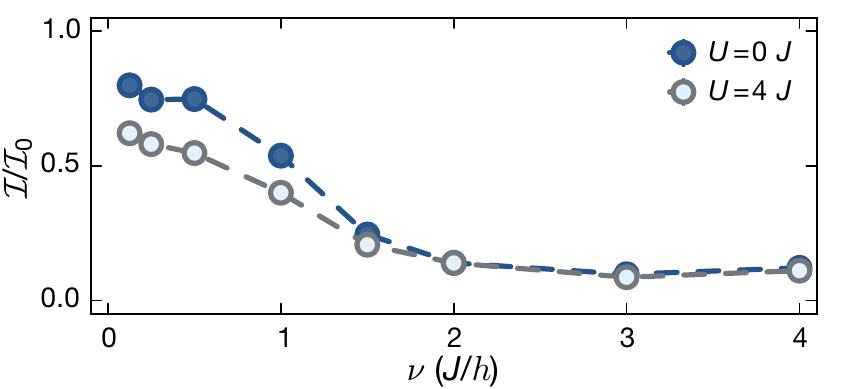}
	\caption{\textbf{Interaction effects on the ratio of the asymptotic imbalance between the driven and undriven systems.} We measure the imbalance for an interacting $U = 4\,J$ and non-interacting $U=0$ system with disorder $\Delta = 7.5\,J$ and drive amplitude $A = 0.7\,\Delta$ as a function of the drive frequency $\nu$. At low drive frequencies the imbalance of the interacting system is stronger suppressed.}
	\label{i_ni_nu}
\end{figure}

\paragraph{\textbf{Interaction effects on the imbalance.---}}We find that at low frequencies  $\nu \lesssim 2\,J/h$ and strong drive amplitude $A=0.7\,\Delta$, the ratio of the saturated imbalance of the driven and undriven systems depends on the interaction strength, even at strong disorder $\Delta = 7.5\,J$, \fig{i_ni_nu}. We attribute this observation to the shift of the critical disorder strength of the MBL transition toward larger values when increasing the interactions. We have checked that in the weak drive limit ($A \ll \Delta$), the saturated imbalance does not display a relevant interaction dependence (data not shown).\\

\end{document}